\documentclass[paper, 12 pt]{article}  
\usepackage[latin1]{inputenc}
\usepackage[lined]{algorithm2e}
\usepackage{amssymb}
\usepackage{graphicx}
\usepackage{multicol}
\usepackage{multirow}
\usepackage{color}
\usepackage{amsmath} 
\usepackage{theorem}
\usepackage{units}

\usepackage{url}
\def\Max{\mathop{\rm max}\nolimits}

\usepackage{caption}

\usepackage{subcaption}

\newtheorem{prob}{Problem}

\title{ \bf
Mathematics and Flamenco: An Unexpected Partnership
\thanks{This work has received funding from the European Union's Horizon 2020 research and innovation programme under the Marie Sk\l{}odowska-Curie grant agreement No 734922. \protect\includegraphics[height=1em]{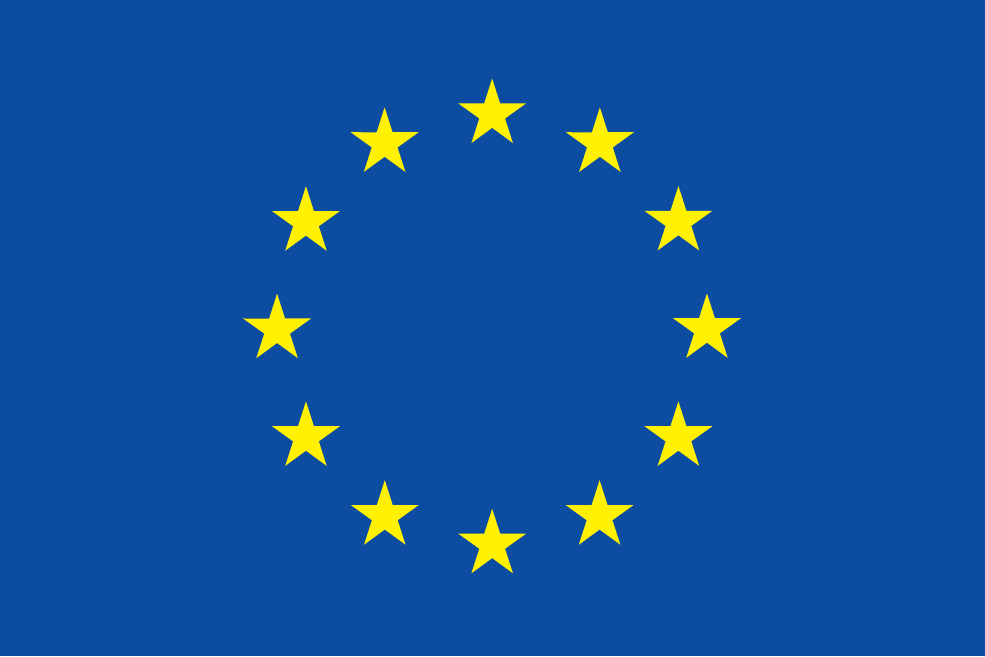}
}
}

\author{J. M. D\'{\i}az-B\'{a}\~{n}ez\thanks{Departamento de Matem\'{a}tica
Aplicada II, Universidad de Sevilla, Spain. 
{\tt dbanez@us.es.}} 
}

\date{}

\begin{document}

\maketitle

\begingroup
 \setlength{\parskip}{6pt plus 1pt minus 0.5pt}
\section{Introduction}
In this paper, we present a series of mathematical problems which throw interesting lights on flamenco music. 
More specifically, these are problems in discrete and computational mathematics suggested by an analytical (not compositional) examination of flamenco ``\emph{cante}'' (singing). As a consequence, since the problems are taken from a culturally specific context, the examples can make more effective mathematics education. There is a vast literature dealing with the relationship between culture and mathematics, much of it concerning the term ethnomathematics \cite{ascher2000,barton1996,presmeg07}.

The questions analyzed here are a selection of problems that arose in the course of the COFLA project\footnote{COFLA: COmputational analysis of FLAmenco music. \url{www.cofla-project.com.}}. Some of them have already been examined, while others are open problems currently being worked on. Their difficulty spans a considerable range, some being suitable for assignments in mathematics classes and others for research projects, depending on the level of mathematical sophistication and knowledge required to tackle them.

Flamenco music was born in Andalusia and is well known around the world. It has been declared an Intangible World Heritage by UNESCO\footnote{\url{http://www.unesco.org/culture/ich/en/RL/flamenco-00363}}. Flamenco music is a unique musical genre with an identity and originality worthy of scientific study. In this article, we pose several problems which arose during a research project which analyzes flamenco rhythms and melodies. We propose several possible answers based on a study that uses mathematics as a codifying operator, focused around three basic questions:
\begin{enumerate}
\item What musical properties are preferred by flamenco aficionados; what do they find most interesting in a piece?
\item How can flamenco rhythms be compared automatically? Or, to put it another way, how can the similarity between two pieces be quantified?
\item How can a piece of flamenco music be simplified or transcribed at a lower level of complexity?
\end{enumerate}

The search for preferred properties in music in the oral tradition leads to the field of ethnomusicology, a scientific discipline formerly called comparative musicology. 
By ``preferred properties'' we mean here the preferences of flamenco aficionados, who are linked specifically to the cultural aspects of the people or region the music belongs to. For more information about ethnomusicology and computational ethnomusicology we recommend the monographs \cite{neetl-83} and \cite{Hol10}, respectively.

The study of musical similarity is part of the fields of music technology, musicology, and psychology. It is related on one hand to the acoustical properties fundamental frequency, harmonics and energy among other aspects, and on the other, to musical properties such as rhythm, harmony and phrasing. It also takes into account psycho-social variables such as emotion, character and sociocultural aspects. Some notable work in this area can be found in \cite{HSF98} and \cite{sch99}.

Lastly, the problem of automatic transcription involves such areas as signal processing, musical psychology, and musical information retrieval (MIR). For more information about automatic transcription of flamenco cante, see \cite{emilia-bonada, nadine}.

In addition, these types of questions are suitable for student assignments and interested readers, both in music and mathematics at various levels. They can be used to broaden the perspective of the general public about mathematics, and also about flamenco, often considered a minor genre by those not familiar with it.

\section{
A short note about flamenco cante}

In this section, we give a brief introduction to the world of flamenco music for readers not previously familiar with it. Being in the oral tradition, flamenco music is created, composed, and transmitted in a different way than Western or classical music. Flamenco has its own rules and its own terminology. Moreover, its standards for evaluating and performing are not necessarily the same as those for other types of music.

Since the problems presented in this paper are derived from a study of \emph{cante} rhythm and melodies, the basic concepts we will treat here relate only to these aspects. For a more detailed description of other elements such as harmony, structure, form, metrics and themes, readers are referred to the manual \cite{gamboa-05}.
Some of the fundamental aspects of flamenco music are \emph{comp\'as} (beat or measure),  ornamentation and improvisation. An essential element of the genre is the seemingly paradoxical blending of ``obligation'' with ``freedom.'' This will be explained in detail but without technical rigor.

\emph{On comp\'as:} The flamenco \emph{cante} (note that this is different from \emph{canto}, a Spanish word for ``singing'') uses a meter (comp\'as) in which the accented beats are distributed in a particular pattern in a short sequence that repeats at regular intervals. The singer must adjust the phrases of the music to this periodic cycle of accents and, in fact, one of the worst criticisms that flamenco aficionados can make of a singer is to say that they ``lag behind,'' meaning that they are not keeping up with the pattern of the comp\'as. When the performer has ``interiorized'' the rhythm, they can play with the timing, using musical resources such as rests and longer phrases that enhance the rhythmic color of the performance. The rhythm is thus a blend of freedom and restriction.

The terms comp\'as (beat, pulse, or musical measure) and rhythm have a specific meaning in flamenco. The comp\'as is the pattern of accents (the sequence of handclaps or taps) that is repeated throughout the whole cante, and can be heard in the guitar and percussion accompaniment. Rhythm is a broader concept, referring to how the cante, toque or baile is performed, which may involve ``playing'' with the timing over the base laid down by the strict meter or comp\'as.

Flamenco rhythms have three types of measures or distributions of accents: binary (a strong accent every 2 or 4 beats), ternary (every 3 beats), and amalgamated 12-beat rhythms. Figures \ref{compases-box} and \ref{notacion} show flamenco ternary rhythmic patterns in box and numerical notation (used in dancing schools) respectively. The strong accents are placed as shown. The reader can play them by clapping louder where there are dots or bold numbers and softer otherwise. Each comp\'as is labelled with the name of the flamenco style or \emph{palo} that uses it, although the compases are not unique to the styles indicated. For example, the fandango comp\'as is also used in \emph{sevillanas} and sometimes in \emph{buler\'ias}. The \emph{sole\'a} comp\'as is also used in \emph{alegr\'ias}, and the \emph{guajira} comp\'as in \emph{peteneras}\footnote{A brief introduction and listening can be found at our website www.cofla-project.com}. 

\begin{figure} [htb]
                        \begin{minipage}[t]{6cm}
                          \begin{center}
            {\includegraphics[width=6cm]{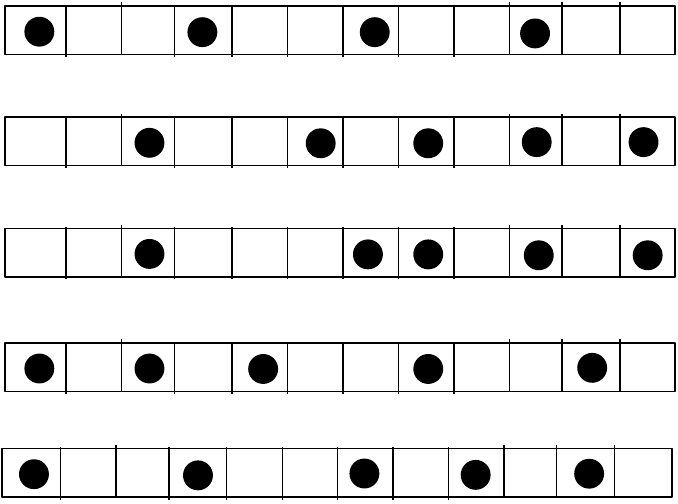}}
            \vspace{0.05cm}
 \caption{
                Box notation. Five 12-beat rhythmic patterns.}
           \label{compases-box}
\end{center}
                        \end{minipage}
                    \hfill
                        \begin{minipage}[t]{6cm}
\vspace{-4.4cm}
 \mbox{[\textbf{\large 1} 2 3 \textbf{\large 4} 5 6 \textbf{\large 7} 8 9 \textbf{\large 10}
11 12]}-\emph{fandango}

\vspace{.1cm} \mbox{[1 2 \textbf{\large 3}  4 5 \textbf{\large 6} 7 {\bf
\large 8} 9 \textbf{\large 10} 11 \textbf{\large 12}]}- 

\emph{sole{\'a}}

\vspace{.02cm} \mbox{[1 2 \textbf{\large 3} 4 5 6 \textbf{\large 7} {\bf
\large 8} 9 \textbf{\large10} 11 \textbf{\large 12}]}-\emph{buler\'{\i}a}

\vspace{.05cm} \mbox{[\textbf{\large 1} 2 \textbf{\large 3} 4 \textbf{\large 5} 6  7
\textbf{\large 8} 9 10 \textbf{\large 11} 12]}-\emph{seguiriya}

\vspace{.05cm} \mbox{[\textbf{\large 1} 2 3 \textbf{\large 4} 5 6 \textbf{\large 7}
8 \textbf{\large 9} 10 \textbf{\large 11} 12]}-\emph{guajira}
    \caption{
                    Numerical notation of 12-beat flamenco rhythmics.}
    \label{notacion}
                        \end{minipage}
                \end{figure}

\emph{On melody:}
Little is known about the origins of flamenco melodies. Some are derived from popular or folk songs that were adapted to flamenco by flamenco singers, while others were composed by the singers themselves. Examples of the former include the \emph{petenera} melody, and of the latter, \emph{personal fandangos} which may be named for the place where they were composed or for the singer or supposed composer. Examples include the \emph{fandango de Huelva, de Alosno, fandango del Ni\~no Gloria, del Carbonerillo}, and \emph{fandango de Chocolate}, among others.

A notable feature of flamenco melodies is the abundant use of ornamentation, melisma (multiple notes sung on a single syllable) and the apparent lack of a steady rhythm. The last feature is due to the fact that the melodic rhythm of the cante generally does not strictly follow the meter or comp\'as played by the percussion and guitar, although the melodic phrase must synchronize with the comp\'as at the end of each cycle. Moreover, although there is an established melodic pattern for each cante,  the melodies are performed with variations on the base pattern that depend on the abilities and esthetic preferences of the singer or the school of cante to which the performer adheres. The same cante with the same words will sound differently performed by, say, Antonio Mairena than by Chano Lobato\footnote{Two audios can be listen at www.cofla-project.com/introduction.html.}. This is due in large part to their individual use of phrasing (organization of the phrases), ornamentations, and timbre, which give flamenco cante a readily recognizable esthetic that also distinguishes it from other types of songs. In any case, flamenco music has certain types of ornamental motifs and phrasing in common that make it an interesting subject of musicological study. Lastly, from the point of view of music technology (the study of music by means of computers), the elaborate ornamentation of flamenco makes it difficult to automate the separation of the notes of the melodic pattern from the melismatic ornamentation \cite{granada, ismir2012-emilia,mora}. An example can be displayed on the videos https://www.youtube.com/watch?v=YpeeOp0Picw and https://www.youtube.com/watch?v=ZGykopQfD2Y

\section{Geometric representation of rhythm and me\-lo\-dy}

\subsection{Comp\'as}

In order to study flamenco music mathematically, we need to codify the musical parameters that we wish to analyze. In the previous section, we saw two different  notations for rhythmic patterns; numeric and box notations. We will now show two geometric notations (transcriptions), chronotonic and polygonal representations.

 Chronotonic or histogram representation was first proposed for automatic voice recognition \cite{gustafson-88}. Let us consider the rhythmic pattern of the seguiriya, which is giv\-en by $\mbox{[\textbf{1} 2 \textbf{3} 4 \textbf{5} 6  7 \textbf{8} 9 10 \textbf{11} 12]}$. This numerical representation does not indicate the relative durations between two consecutive beats. The time between one accent (musical attack or onset) and the next defines the dynamics of the rhythm.
For a visual representation that conserves information about time, we depict the time between every pair of accents in two dimensions as shown in Figure \ref{crono}. Note that the time between every pair of accents is shown in two dimensions (horizontal and vertical), and the representation starts at the beginning of the rhythmic pattern. Observe that Sole\'a and Seguiriya share in common the absence of accent on the first beat (\emph{anacrusis}) and then the number of steps increases in one with respect to the other representations.
Each duration between accents (rhythmic interval) is represented by a two-dimensional box, and both axes, $x$ and $y$, represent the length of time of an interval. The representations in Figure crono can be seen as step functions meaning the temporal profile or behavior of the comp\'as with respect to the ``waiting time'' between two consecutive attacks.

\begin{figure} 
                          \begin{center}
            \includegraphics[width=10cm]{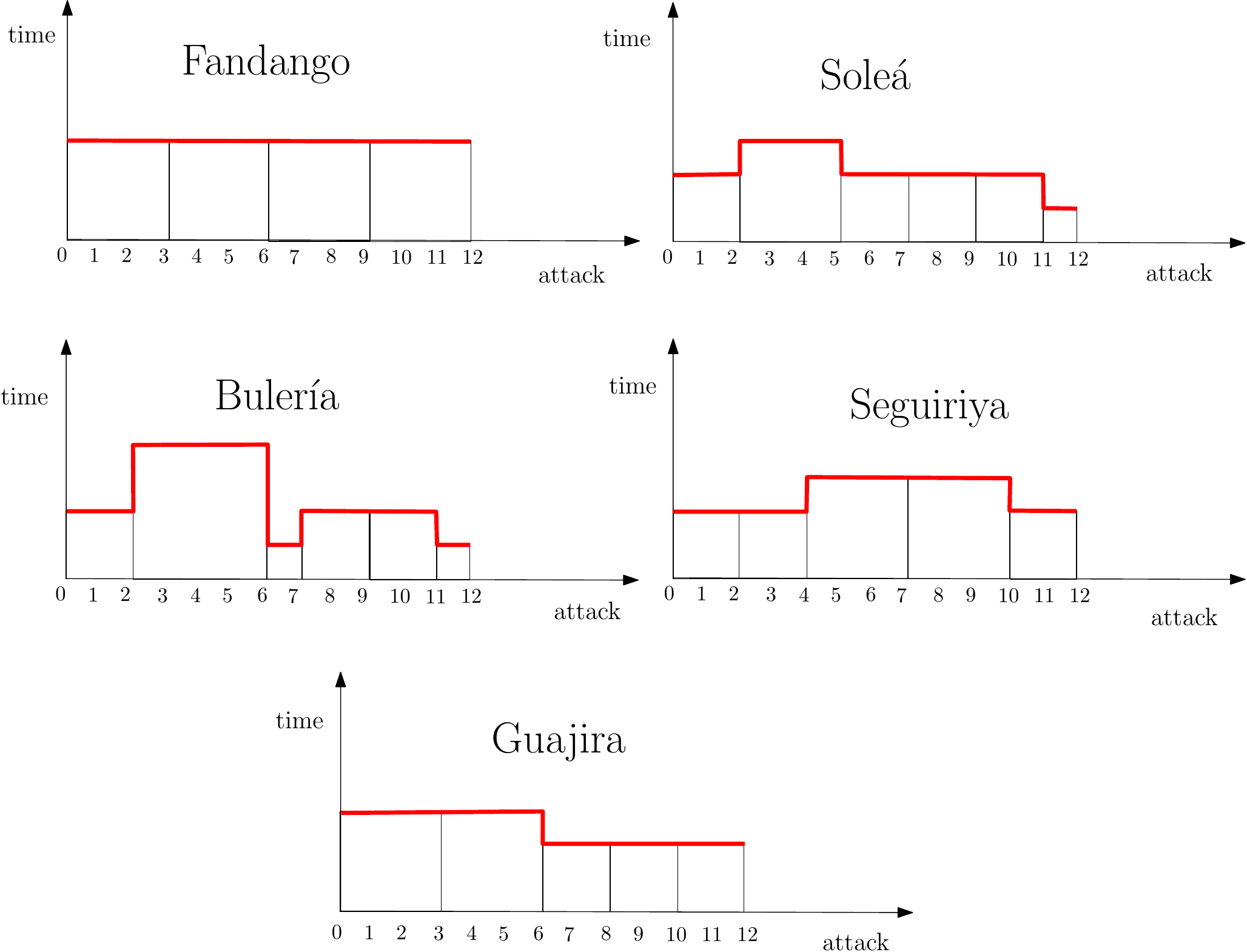}
\vspace*{0.5cm}  \caption{Chronotonic representation of ternary flamenco patterns.}
           \label{crono}
\end{center}
\end{figure}

The polygonal representation  depicts the twelve beats of the comp\'as as equidistant points arranged around a circle. We could imagine it as a necklace of white (weak) and black (strong) beads  (accents) arranged like the numbers on a clock (Fig\-ure \ref{poligonos}).  If we join each pair of successive strong accents, we draw a polygon whose vertices are exactly the black beads. In the polygonal representation shown in Fig\-ure \ref{poligonos}, the ``1'' indicates the position where the rhythmic pattern begins, and the vertices are the strong accents. Note that the chronotonic codification starts at the ``1" indicated in Figure \ref{poligonos} and takes into account the times to the next event.
A notation which represented the notes of a musical scale by a polygon in a clock diagram appeared as early as 1937 in an article by E.~Krenek \cite{krenet-02}. We apply it here to visually depict flamenco comp\'as.

\begin{center}
\begin{figure}[htb]
            \centerline{\includegraphics[width=10cm]{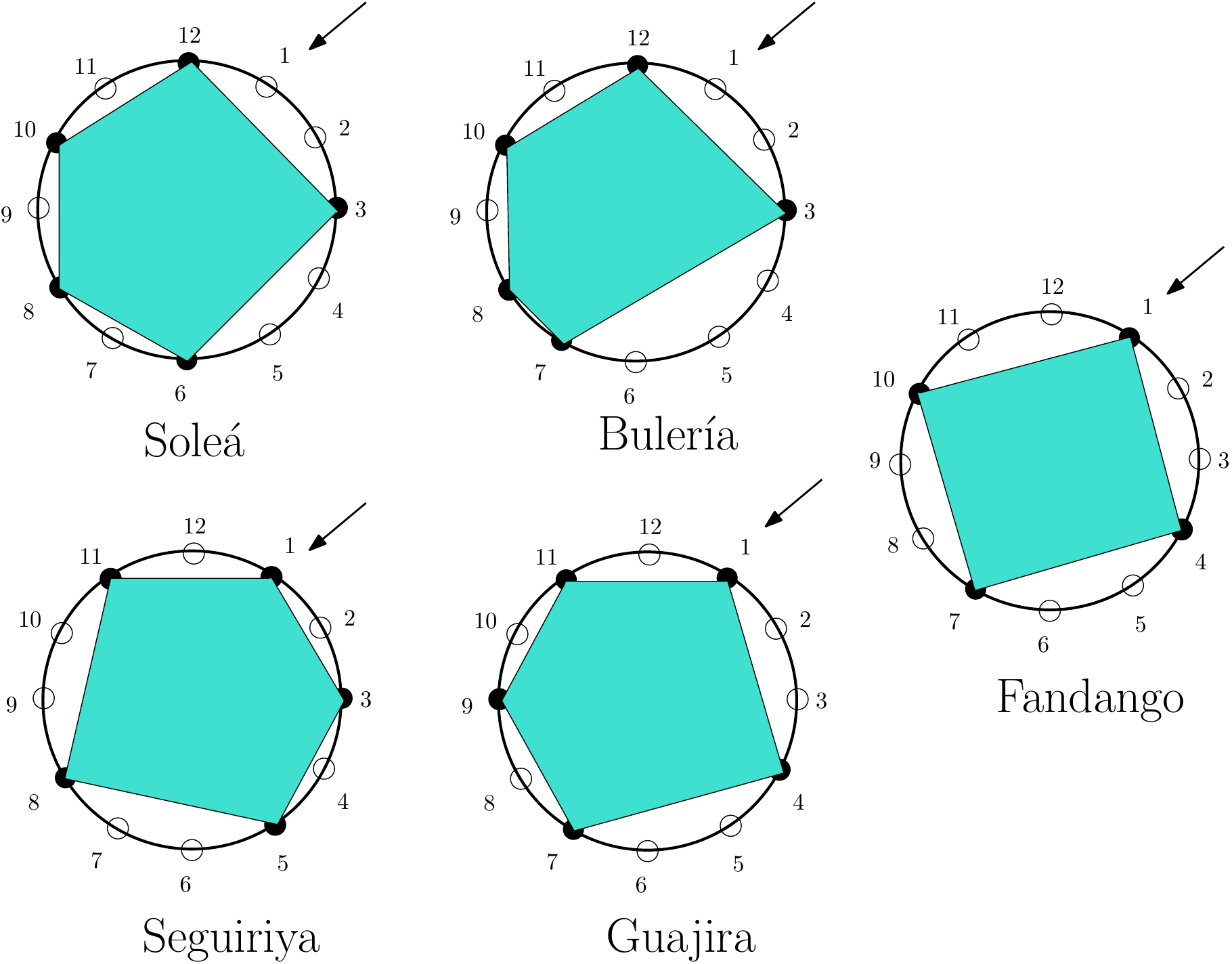}}
\vspace*{0.5cm} \caption{Five ternary flamenco compases.}
           \label{poligonos}
        \end{figure}
\end{center}

\subsection{Melody}
\label{melod}

Melody is the parameter that defines and distinguishes a particular song. In fact, the word ``melody'' comes from the Greek word \emph{meloid\'ia} ($\mu\epsilon\lambda\omega\delta\text{\'{\i}}\alpha$) which means ``choral song'' and is identified with the song itself.\footnote{Interestingly, in the mariachi music of Mexico, the word ``\emph{melod\'ia}'' (``melody'') is used in a similar way to mean a song.}

A melody can be represented symbolically (a string of characters) or by audio (a frequency curve or amplitudes from the audio signal). Common symbolic representations are score transcriptions and MIDI files. We term such symbolic formats discrete (a finite amount of data), and audio representations continuous (since they are given by a frequency curve), respectively. Since discrete representations lend themselves better to musical analysis, various different ``simplifications'' of melodies have been proposed, depending on the analysis to be carried out. For example, the same level of detail in the transcription of a melody is not required to distinguish between two songs as between two performances of the same song. The simplest symbolic representation of a melody is a string of alphabetic symbols corresponding to the notes; for example, ``CDE.''

A melody can be defined discretely as a series of sounds with particular pitches and durations; that is, a series of sound frequencies with an associated rhythm that is perceived as single identity. By this definition, a melody can have a simplified representation as a succession of pairs $M=\{(t_1,f_1), (t_2,f_2), \ldots, (t_n,f_n)\}$, $t_1<t_2<\cdots<t_n$, where $n$ is identified as the length of the melody. Since this representation is not invariant to changes in pitches or tempo, a representation by intervals is also used; that is, $I_1=(t_2-t_1,f_2-f_1), I_2=(t_3-t_2,f_3-f_2), \ldots, I_{n-1}=(t_n-t_{n-1},f_n-f_{n-1})$ and so $M=\{I_1,I_2,\ldots,I_{n-1}\}$. The importance of the representation by intervals is that the  melodic direction matters for some analyses. If we joint the points of the discrete set $M=\{(t_i,f_i)\}$, we obtain a polygonal curve termed the \emph{melodic contour}.  Note that the peaks or extreme points of the polygonal function (piecewise linear) are a key characteristic of the melody. Let us examine the monophonic flamenco cante, the \emph{debla} ``\emph{En el barrio de Triana}'' as an example. Figure \ref{smstools-debla} shows the audio signal (upper diagram) and the fundamental frequency (lower diagram). The lower part shows a segmentation (simplification of the melody). The data for this segmentation in the time interval $[0,8]$; that is, the first phrase of the melody, are M=\{(0.2,\ 385), (0.4,\ 407), (2.2,\ 407), (3,\ 385), (3.3,\ 407), (4.3,\ 385), (4.7,\  407), (5.2,\  385),  (5.8,\ 407), (6.1,\  385), (6.5,\ 330)\}. In interval notation, each two-dimensional point is of the form $(\Delta(t_i),\Delta(f_i))$, and we can draw the melodic profile by joining the points.

In summary, a melody can be represented in a variety of ways depending on the information we wish to extract; by a continuous function, a step function (lower part of Figure \ref{smstools-debla}) or polygonal function that is monotone respect to the $X$-coordinate.

\begin{center}
\begin{figure}[htb]
            \centerline{\includegraphics[width=12cm]{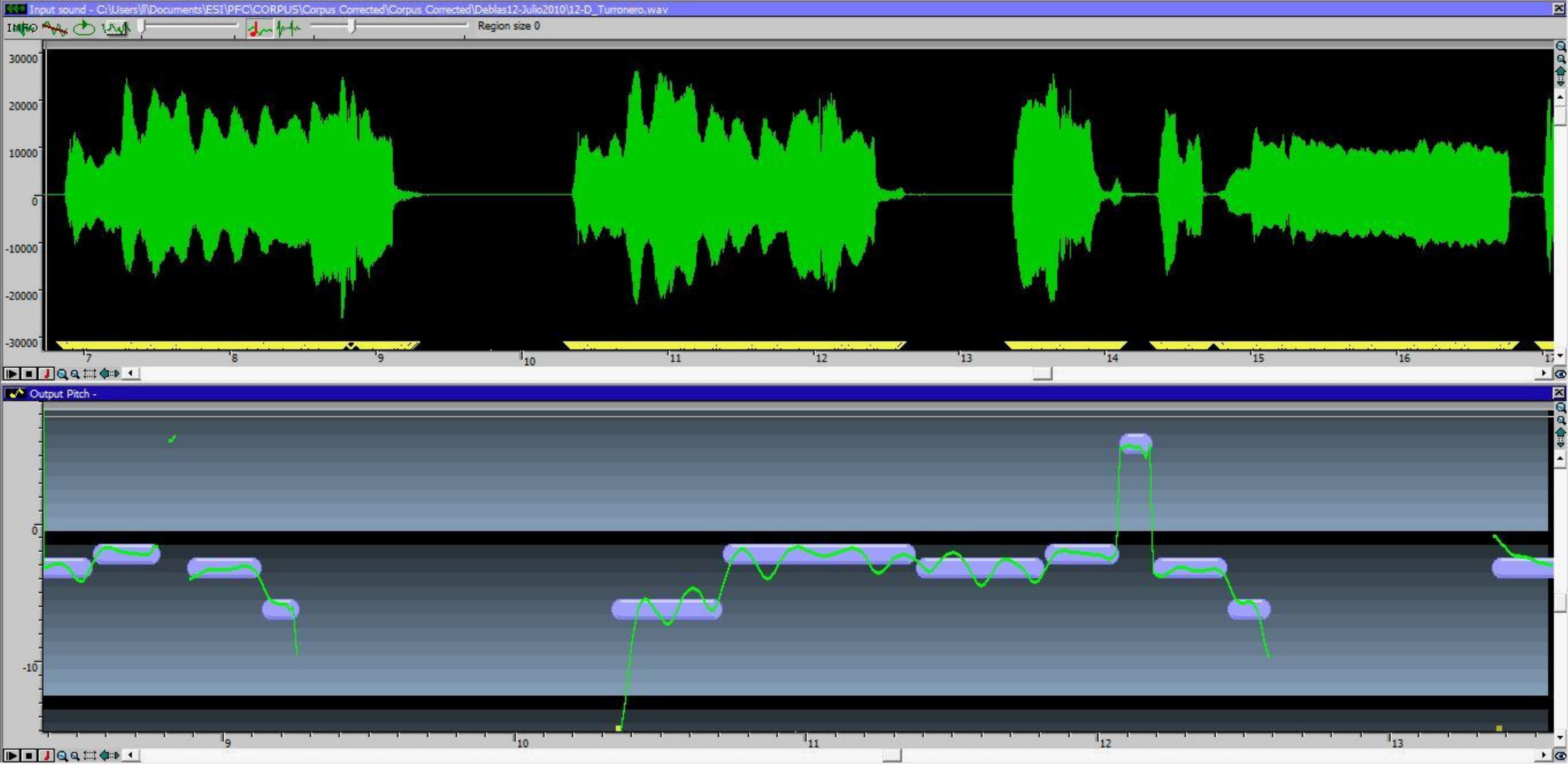}}
\caption{
                  Automatic transcription of a phrase of a flamenco melody.}
           \label{smstools-debla}
        \end{figure}
\end{center}

\section{
              Regular rhythms: Optimization problems}
\subsection{
                    Maximum area rhythms}

One of the properties that musicologists and mathematicians have observed in music in various oral traditions is what is called regularity of the rhythm (or of the musical scale) \cite{godfried-2010}. Regular rhythms have been defined as those which maximize a particular geometric measure. For example, in the polygonal representation of a rhythm, we can use as measures of regularity the sum of the distances between each pair of strong accents (black beads in the necklace) or the area of the rhythm polygon. This suggests the following question:

\begin{prob}[\textbf{Inscribed pentagon of maximum regularity}] Given so\-me measure of regularity (e.g., area, sum, or other) and 12 points uniformly distributed around a circle, what choice of five of these points will maximize regularity by the specified measure? \label{prob:regpolyg}
\end{prob}

Problems of this type involve areas of mathematics such as combinatorics and geometry, and have been studied by mathematicians for some time.
In 1956, the Hungarian mathematician F.~T\'oth \cite{toth-56} examined the continuous case. He proved that the set of $n$ points on the circle that maximizes the sum of the distances between each pair of consecutive points is precisely the vertices of a regular polygon with $n$ sides. To put it another way, the vertices are distributed as regularly as possible in the circumference of the circle. If the criterion of maximum area is considered, it can be proved using basic geometry that for $n=3$ (three points), a triangle inscribed in a circle has maximum area when it is regular; that is, an equilateral triangle\footnote{This can be proved formally by a \emph{reductio ad absurdum} argument in high school.}. The general case; that is: The $n$-vertex polygon of maximum area inscribed in a circle is the regular $n$-gon, can be proved using geometric arguments. The proof will be omitted here.

In applying this problem to music, the case of interest is the discrete version of the problem; that is, choosing the vertices of the polygon drawn on a set of fixed points (Problem \ref{prob:regpolyg}). The general discrete problem is:
\begin{prob}[\textbf{Inscribed polygon of maximum area}] Consider a circle with $n$ points located evenly around its circumference. How should $k<n$ of the points be selected in such a way that the polygon whose vertices are the $k$ points will have maximum area of all such possible $k$-gons?
\end{prob}

We consider here only the case relevant to flamenco music, $k = 5$ and $n=12$, although the
arguments can easily be generalized.
The proof that the polygon corresponding to the sole\'a comp\'as (and consequently, seguiriya and guajira as well) is the pentagon of maximum area is as follows. Since the 12 points are distributed uniformly around the circumference of the unit circle, the area of any inscribed polygon based on the points can be written as the sum of triangles anchored to the center $O$ of the circle (Figure \ref{max-area}). These triangles are isosceles because the radius of the circle forms two of their sides. The interior angle of the vertex in the center of the circle is, in fact, 30 degrees. Thus the distribution of the 5 black points chosen from the set of 12 can be encoded as a vector giving the number of triangles between each pair of consecutive accents. For example, $33222$ is a sole\'a and $34122$ is a buler\'ia. To see that the sole\'a has a greater area than the buler\'ia, it is sufficient to observe the inequality between the marked subpolygons in Figure \ref{max-area} (since the area of the rest of the polygon is the same for both cases). We note that although it is visually evident, a formal proof is still needed, since appearances can be deceiving. The formal proof uses basic trigonometry. It is easy to prove that the area of triangle $O78$ (the triangle whose vertices are the center $O$ and points 7 and 8) is exactly half  the sine of $30$ degrees; that is,  $A_{O78}=\frac{\sin 30^{\circ}}{2}$. Similarly, the area of triangle $O68$ is  $A_{O68}=\frac{\sin 60^{\circ}}{2}$. Thus the area of the polygon drawn for the sole\'a is
\[A_{O68}+A_{036}=\frac{\sin 60^{\circ}}{2}+\frac{\sin 90^{\circ}}{2}=\frac{\sqrt{3}}{4}+\frac{1}{2},\]
and the polygon for the buler\'ia
\[A_{O78}+A_{O37}=\frac{\sin 30^{\circ}}{2}+\frac{\sin 120^{\circ}}{2}=\frac{1}{4}+\frac{\sqrt{3}}{4}.\]

\begin{figure}[htb]
    \begin{center}
        \includegraphics[width=10cm]{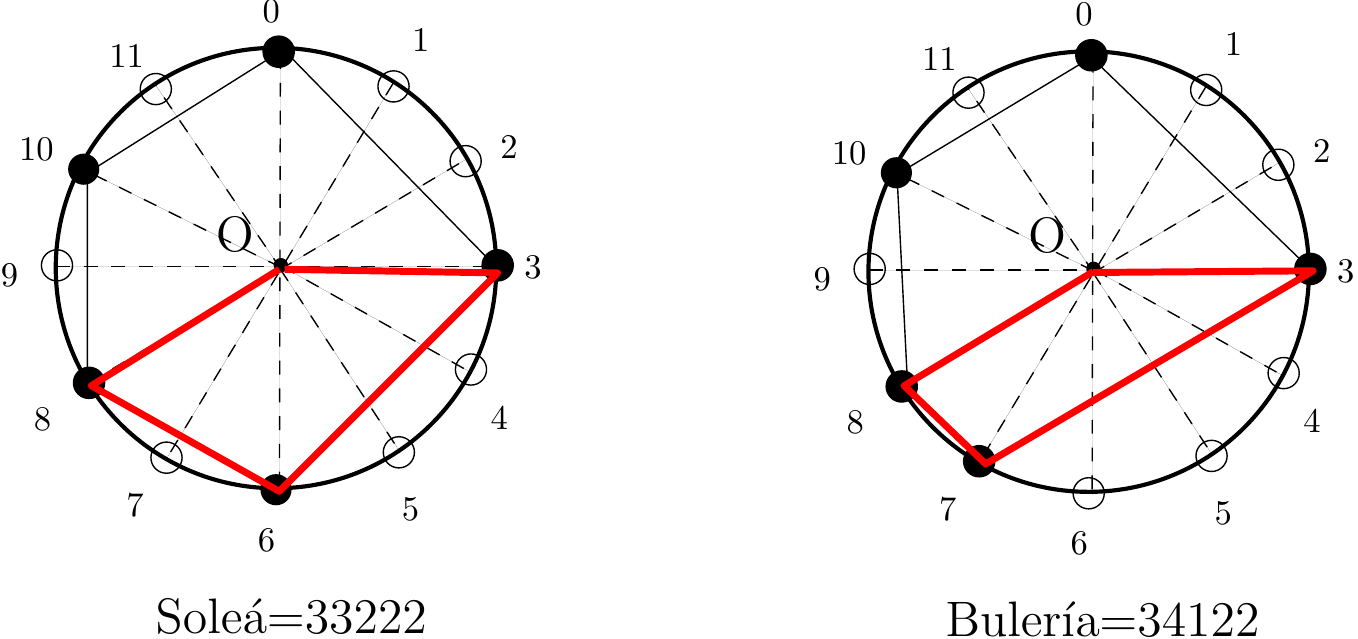}
        \caption{\small{
                                Sole\'a has greater area than buler\'ia}.}
        \label{max-area}
    \end{center}
\end{figure}

This proves that the area of the polygon representing the sole\'a is greater than the area representing the buler\'ia. In fact, it is not difficult to prove that 33222 is associated with the greatest possible area up to permutations (the area of 32322 is the same as that of 33222 because it is only a question of permuting two triangles). Note that a distribution different from one that contains two 3's and three 2's must contain one 4 and one 1. If the triangles corresponding to the 4 and the 1 are placed next to each other, they can be transformed into a ``3'' triangle and a ``2'' triangle, which, as we have seen, add up to a greater area. As a consequence, the 33222 and 32322 are the distributions of five points with the greatest area. The sole\'a, seguiriya and guajira rhythms all correspond to 33222; the difference between them is where on the circle the comp\'as starts.
The sole\'a, seguiriya and guajira have the maximum area, but the buler\'ia does not. The fandango rhythm also corresponds to the maximum area over all choices of four points out of twelve equidistant points arranged around the circumference.

In conclusion, turning to the musicological application of this finding, most flamenco rhythms have the ``maximum area'' property, and the buler\'ia\footnote{It is interesting to note that this rhythmic meter, known as the ``modern \emph{soniquete}'', is basically sung using the \emph{buler\'ia por sole\'a}, a hybrid of the buler\'ia and sole\'a.} does not have the regularity of the others. Further research could test human subjects (flamenco aficionados and others) to confirm or refute the theory that the maximum area property characterizes preferred flamenco rhythms. 

Although an etnomusicological study is beyond the scope of this paper, we include two comments that can be inferred from the study above.
First,  although the origin of the flamenco rhythmic patterns  has not yet been established one can think that the oldest one is the 4/12 
(3/4) pattern used in fandangos and the other compound patterns were derived form the 4/12 while maintaining the regularity property. However, in \cite{d-bfgrt-04} it is claimed that guajira pattern seems to be the ``genetic forbear'' of the 5/12 flamenco rhythmic patterns. See also the subsection 5.2.
Secondly, the pattern used in buler\'ia (and buler\'ia por sole\'a, alegr\'ia, etc.) can be interpreted as a syncopation phenomenon on the sole\'a pattern. The recent emergence of this pattern in flamenco recordings also supports that idea. 
Anyway, these and other insights suggested from a mathematical study poses a possible ethnomusicology research project to confirm these theories. For a description of preferred properties in flamenco music, see \cite{db-itamar}.

\subsection{
                       Minimizing the maximum rest: The minmax problem}

Let us now examine the maximum area problem as a geometric problem of approximating polygons. Given the regular polygon whose vertices are the 12 points of the clock diagram, we can see how it differs from the sole\'a polygon. Figure \ref{oreja} shows the difference in area between these two polygons (the regular and the sole\'a polygons). The difference in area can be expressed as a sum of the areas between each pair of vertices of the sole\'a. We call ``ears'' to the polygons given these areas. Note that maximizing the area of a polygon with 5 vertices is equivalent to minimizing the sum of the areas of its ears. If we denote ear polygons by their end vertices, in Figure \ref{oreja} this means minimizing the area $A(0,3)+A(3,6)+A(6,8) +A(8,10)+A(10,0)$. We call this criterion \emph{minsum} because it seeks to minimize the sum of the areas that are left outside the polygon.

Let us now consider another criterion (as we often do in geometrical optimization) known as the \emph{minmax} criterion. Its goal is to minimize the maximum; that is, to minimize the area of the largest ear. Musically, this can be interpreted as having the shortest possible rest, or time between two consecutive notes. Clearly the buler\'ia rhythmic pattern does not meet the minmax criterion because its largest ear has 5 vertices, and there are other rhythms whose largest ear has a smaller area -- the sole\'a, seguiriya and guajira. The mathematical problem in this section is the following.

\begin{prob}[\textbf{Largest minimum rest}]
Consider a circle with $n$ points dispersed evenly around its circumference. How can $k<n$ of these points be chosen so that the polygon drawn on these $k$ vertices minimizes the area of the largest ear from among all possible choices of $k$ points?
\end{prob}

\begin{figure}[htb]
    \begin{center}
        \includegraphics[width=10cm]{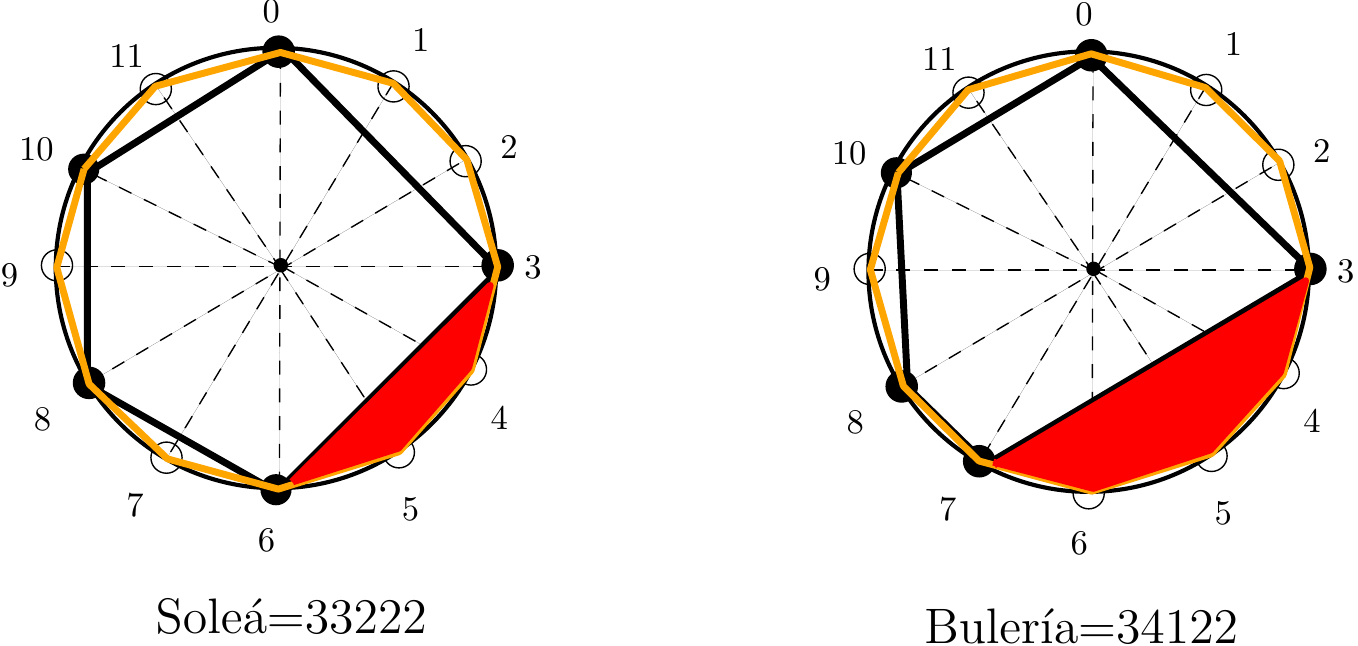}
        \caption{\small{
                                 The sole\'a has a smaller maximum rest (ear) than the buler\'ia.}}
        \label{oreja}
    \end{center}
\end{figure}

It is clear that the sole\'a polygon (also but also guajira and seguiriya) is the solution to the problem for $n=12$ and $k=5$, since there can be no configuration of 5 vertices with all ears having less than 4 vertices. In general, for any $n$ and $k$, it is easy to see how to find the polygons that satisfy the minmax criterion. We  show it here for the case of $n=12$ and $k=5$, and leave the general problem to the reader.

Given a polygon $Q$ with $k(=5)$ vertices inscribed in a circle on the $n(=12)$ given points, we say that $Q$ has length ${\cal L}$ if the longest substring of points between two consecutive vertices of $Q$ has exactly ${\cal L}+1$ vertices. In our example, the sole\'a has length 3 and the buler\'ia has length 4. In fact, the length is the same as one less than the number of vertices (we do not count the last vertex). With $12$ points, $5$ substrings (ears) and the remainder theorem, we have $n=k\cdot q+r$ with $0\leq r \leq k-1$, and, in this case, $12=5\cdot 2+2$. The pigeonhole principle (Dirichlet principle), says that if we have $n$ pigeons to fit into $k<n$ boxes, at least one box must hold $\lceil \frac{n}{k} \rceil$ pigeons. To apply this in our case, we take the black beads as boxes and all the beads as pigeons. Then at least one black bead must be followed by at least two white beads. To state it another way, there must be an ear of length 3, and so any combination of five vertices for which 3 is the maximum length is an optimal polygon; that is, a polygon that minimizes the maximum ear. Thus any distribution of 5 points with two 3's and three 2's will work; for example, 33222 (sole\'a) or 32322 (not a flamenco rhythm).

\section{
               Mathematical measures of similarity: Comparing two flamenco rhythms}

Musical similarity becomes a crucial problem to investigate the origin and evolution of music in an oral tradition. It is also useful for classification of musical styles. The analysis of similarity is an exciting area of research where we can find many interesting mathematical problems. Degree of similarity is calculated using a measure of the distance between two musical interpretations, and the concept of distance between two objects is amply formalized in mathematics. In practice, the calculation of similarity is a difficult problem, since subjective elements such as perception and socio-cultural context come into play. To have an idea of the importance of applications of this concept outside of mathematics, we cite the example of its use in the resolution of legal copyright conflicts in the United States \cite{cronin-98}.
Although mathematical measures of similarity have been proposed, it should be noted that theoretical definitions are not enough; perceptual validation is also required; that is, experiments with groups of people that show that the measures correspond to listeners' perception of similarity. One study of mathematical measures of similarity of rhythms can be found in \cite{hofmann-engl-02,paco}.

We describe two measures of distance from \cite{d-bfgrt-04,db05} which give good results when applied to flamenco comp\'as; \emph{chronotonic distance} and \emph{permutation distance}.

\subsection{
                    Calculating area: Chronotonic distance}

Chronotonic distance was proposed by Gustafson \cite{gustafson-87} in the area of phonetics to measure the similarity between two speech recordings (voice recognition). If we have two rhythmic patterns represented in chronotonic form as in Figure~\ref{crono}, the chronotonic distance is defined as the area of the difference between the two curves. Let us look at an example: if we superimpose the chronotonic representations of the fandango and the seguiriya, the area between them is 6 (see Figure~\ref{distancia-crono}). The chronotonic distance or dissimilarity between any two flamenco rhythms, $d_c$, is calculated in this way, and Table \ref{tabla-crono} gives information on the dissimilarity  between different pairs of flamenco compases using this measure.

\begin{figure} [htb]
                          \begin{center}
            \includegraphics[width=10cm]{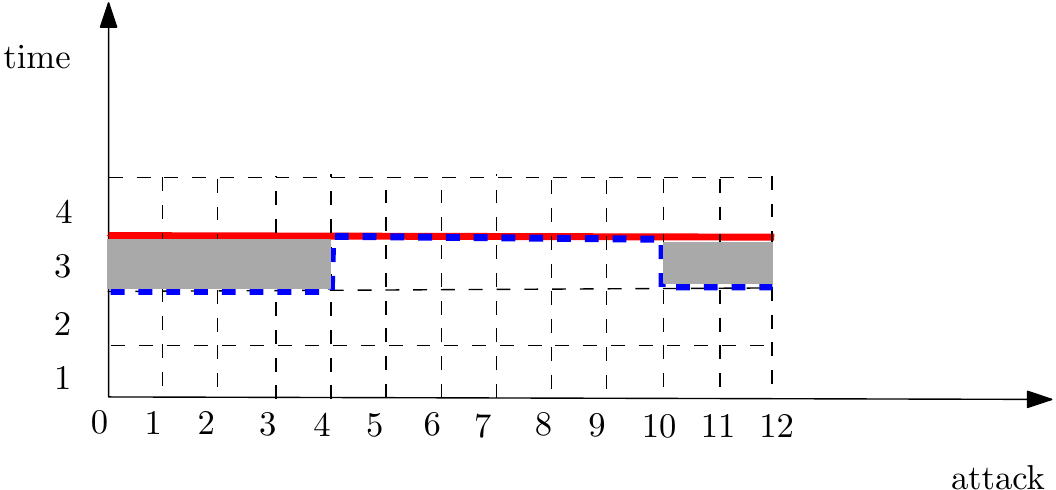}
 \caption{
               Chronotonic distance between fandango and seguiriya.}
           \label{distancia-crono}
\end{center}
\end{figure}

%
%

\begin{table}[h]
\begin{center}
\begin{tabular}{|l|c|c|c|c|c|}
\hline
$d_c$ & Sole\'a & Buler\'ia & Seguiriya & Guajira & Fandango \\
\hline
\multirow{1}{12mm} {Sole\'a} &  0 & 6  & 8 & 4 & 10  \\
\hline
\multirow{1}{12mm} {Buler\'ia} &  6 & 0  & 12 & 8 & 14  \\
\hline
\multirow{1}{15mm} {Seguiriya} &  8 & 4  & 0 & 8 & 6  \\
\hline
\multirow{1}{12mm} {Guajira} &  4 & 8  & 8 & 0 & 6  \\
\hline
\multirow{1}{17mm} {Fandango} &  10 & 14  & 6 & 6 & 0  \\
\hline
\hline
\multirow{1}{12mm} {$\sum$} &  28 & \textbf{40}  & 34 & \textbf{26} & 36  \\
\hline
\multirow{1}{12mm} {\emph{Max}} &  10 & 14  & 12 & \textbf{8} & 14  \\
\hline
\end{tabular}
\caption{
              Table of chronotonic distances.}
\label{tabla-crono}
\end{center}
\end{table}

If we add up the columns in the similarity matrix, we obtain the distance of each respective rhythmic pattern from the others. This quantity, labeled $\sum$ in the table, gives us a numeric value for the ``nearness'' or ``distance'' of one rhythmic distribution from the others.
%
%
%
%
%

By this means, using the criterion of the sum of distances, the buler\'ia rhythm can be said to be ``farther'' from the others, and the guijira comp\'as ``closer.''

Another criterion, instead of the sum of distances, is the maximum, labeled \emph{Max} in Table \ref{tabla-crono}.
%
One possible musical interpretation of this method is that the guajira is a sort of \emph{prototype} representing the identity of flamenco rhythms. However, this is merely a preliminary study of rhythmic patterns; 
a full musical study supported by historical and cultural data would be required for such a conclusion.

\subsection{
                    Calculating exchanges: Permutation distance}

A permutation can involve two terms (not necessarily consecutive) in a series. If two consecutive elements are permuted, we call it a swap or exchange. Note that the pattern of the buler\'ia differs from that of the sole\'a by a swap between elements 6 and 7 (Figure \ref{intercambio}).
\begin{center}
\begin{figure}[htb]
            \centerline{\includegraphics[width=9cm]{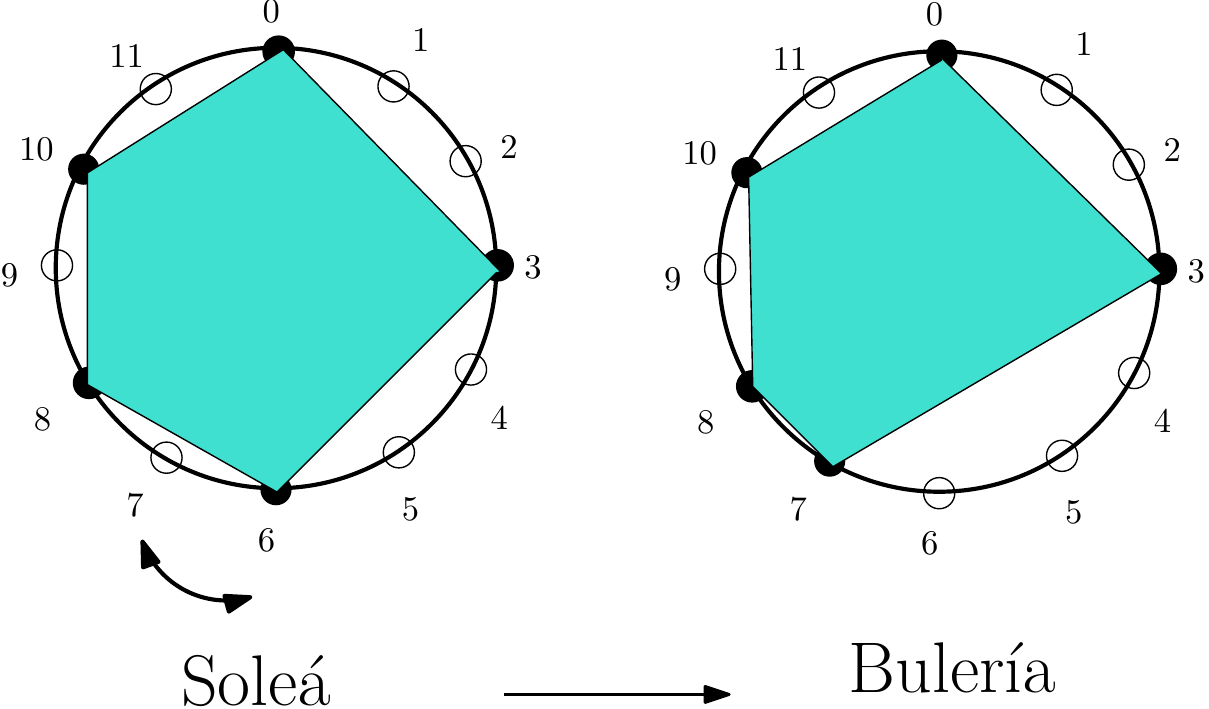}}
\caption{
               One swap is sufficient to change to another comp\'as.}
           \label{intercambio}
        \end{figure}
\end{center}

This small operation produces a significant change in the musical perception of the piece, so it  not surprising that the two patterns are considered different flamenco styles.
We will use precisely this operation (swap) to study the mechanisms of rhythmic similarity. It should be noted that the  ``modern soniquete,'' the pattern which we have labeled buler\'ia (which is common in \emph{buler\'ia por sole\'a}), has a particular artistic effect, since the swap creates a rest from positions 3 to 7, followed by a state of tension with the two consecutive strong accents on 7 and 8. This distribution of the accents (percussive phrasing) gives the handclaps a sequence of tension and resolution that gives the cante, toque, and baile a unique ``flamenco flavor.'' These types of changes are found in the compositions of innovators such as Paco de Luc\'ia (Figure \ref{camaron}) and other flamenco guitarists. 

This discussion has laid the grounds for the definition of permutation distance proposed in \cite{d-bfgrt-04, db05}. 
The permutation distance $d_p(P_1,P_2)$ between two rhythmic patterns $P_1$ and $P_2$ is defined as the minimum number of swaps needed to transform one comp\'as into another. The concept is in the spirit of the \emph{Hamming distance}, proposed to solve problems in applications of \emph{information theory} \cite{hamming-86}. R.~W.\ Hamming (1915--1998) was an American mathematician whose work had many implications for advances in computer science and telecommunications\footnote{During World War II, R.~W.\ Hamming joined the Manhattan Project to solve some equations posed by physicists working on the project. The objective was to determine whether detonating the atomic bomb could ignite the atmosphere. The results of the calculations showed that this would not happen, paving the way for the United States to use the bomb, first in tests in New Mexico and shortly after in Hiroshima and Nagasaki.} (Figure \ref{hamming}).

\begin{figure} [htb]
                        \begin{minipage}[t]{4.7cm}
                          \begin{center}
            \fbox{\includegraphics[width=4.7cm]{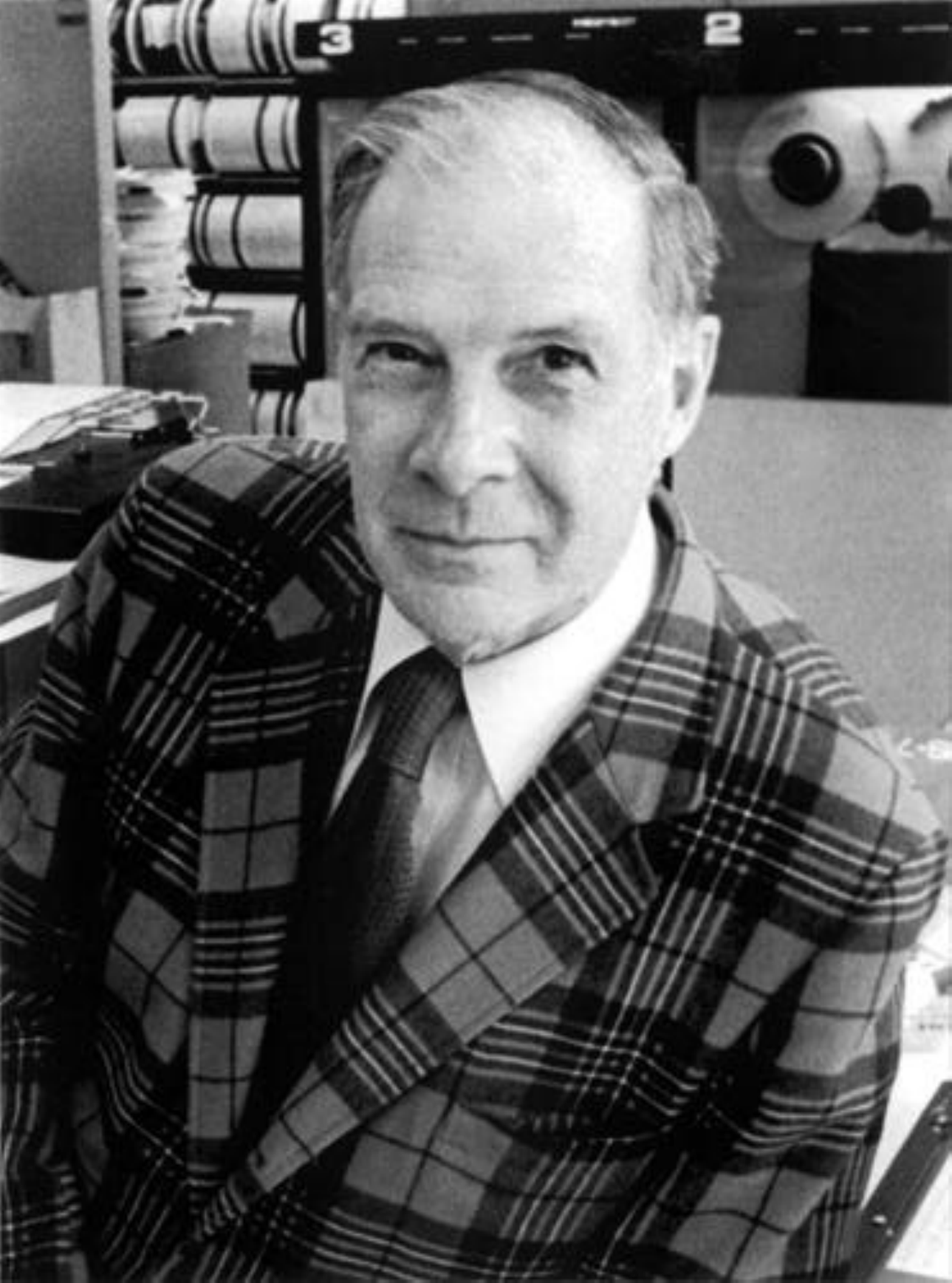}}
 \caption{R.\ W.\ Hamming.}
           \label{hamming}
\end{center}
                        \end{minipage}
                    \hfill
                        \begin{minipage}[t]{7cm}
\vspace{0.95cm}
\fbox{\includegraphics[width=7cm]{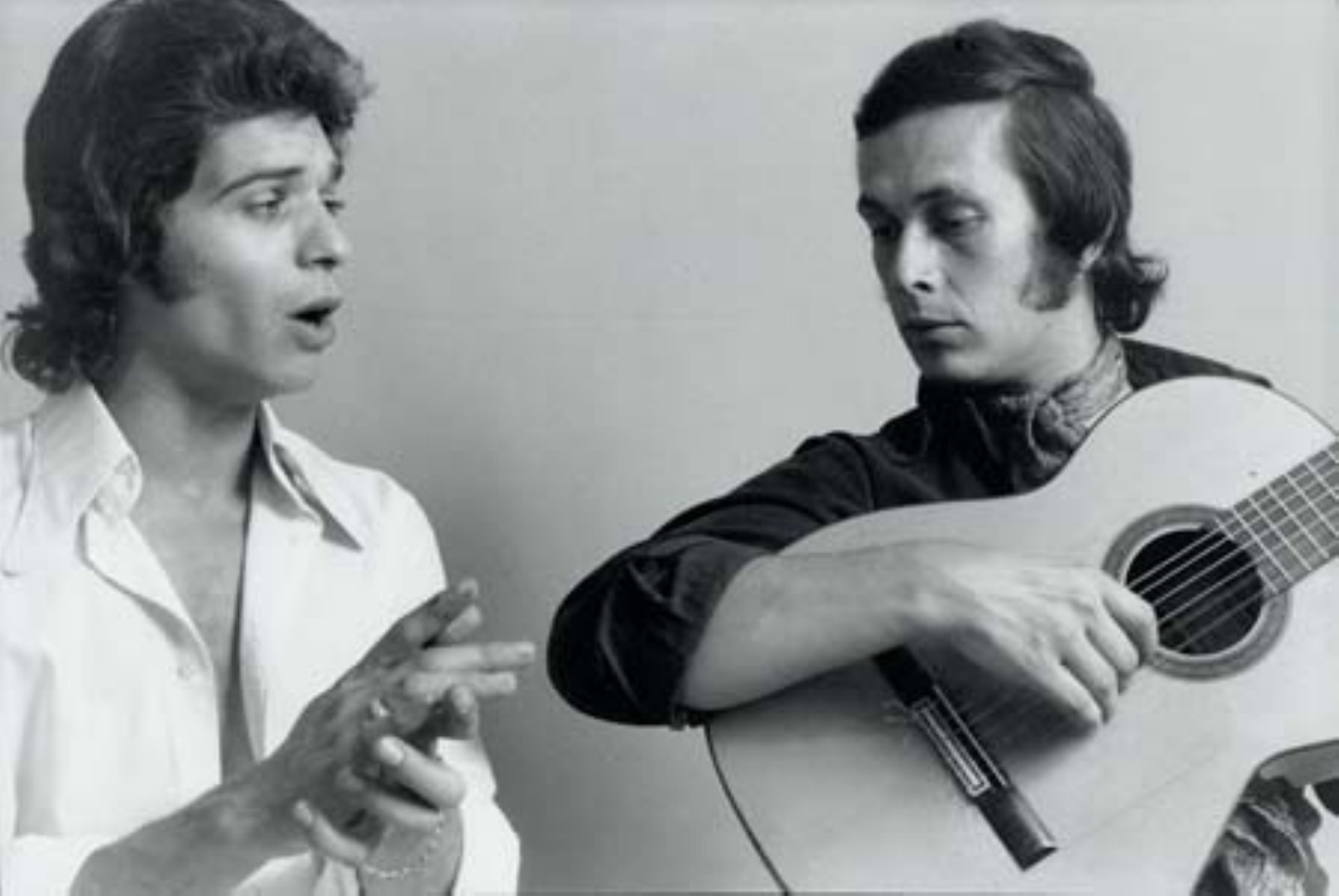}}
 \vspace{0.1cm}
    \caption{
                  Camar\'on de la Isla and Paco de Luc\'ia.}
    \label{camaron}
                        \end{minipage}
                \end{figure}

The Hamming distance between two strings of codes of the same length is the number of substitutions that would be required to transform one string into the other. Let us consider an example:  if we have the binary strings $C_1=[{10} \textbf{1} {1} \textbf{1} {01}]$ and  $C_2=[{10} \textbf{0}{1} \textbf{0} {01}]$, the Hamming distance between them is $d_h(C_1,C_2)=2$. The basic operation which is the counting unit is substitution. In the context of flamenco music, in \cite{d-bfgrt-04} the basic operation was swap or permutation between consecutive beats, and the permutation distance between two  compases was the minimum number of swaps need to transform one comp\'as into the other. If we represent the sole\'a and buler\'ia patterns with ones and zeros, they can be written as $So=[001001010101]$ and $Bu=[001000110101]$. The Hamming distance between them is 2 and the permutation distance is 1.

: 
Once we have two rhythms in binary representation, a certain skill is required to efficiently calculate the permutation distance between them. Here we will treat only the case where the binary strings (compases) have the same number of strong accents (ones). The flamenco rhythms sole\'a, buler\'ia, seguiriya and guajira all have five strong accents. The general problem we wish to solve is the following.

\begin{prob}[Calculating the permutation distance]
Consider two binary strings of $n$ digits (for arbitrary $n$) with the same number of ones (and therefore the same number of zeros). Design a procedure to calculate the permutation distance between them efficiently.
\end{prob}

Note that when $n$ is small, as in the flamenco, where $n=12$, the permutation distance can be calculated by simple observation. Figure \ref{d_p(S,G)} shows the minimum number of swaps needed to transform the seguiriya comp\'as into the guajira comp\'as. But if we think of the general problem for strings of codes of many elements, for example $n=100$, it is clear that we can not do it by hand. To solve the problem, we use the following idea. Let us represent each string by a vector of the positions of the ones. For example, the seguiriya is represented by $Se=(1,3,5,8,11)$ and the guajira by $Gu=(1,4,7,9,11)$. We now observe that we obtain the minimum number of swaps needed by adding the absolute values of the differences between the elements of these auxiliary vectors. What is then the maximum number of operations required to calculate the permutation distance between two strings of $n$ elements?

\begin{figure} [htb]
                          \begin{center}
            \includegraphics[width=8cm]{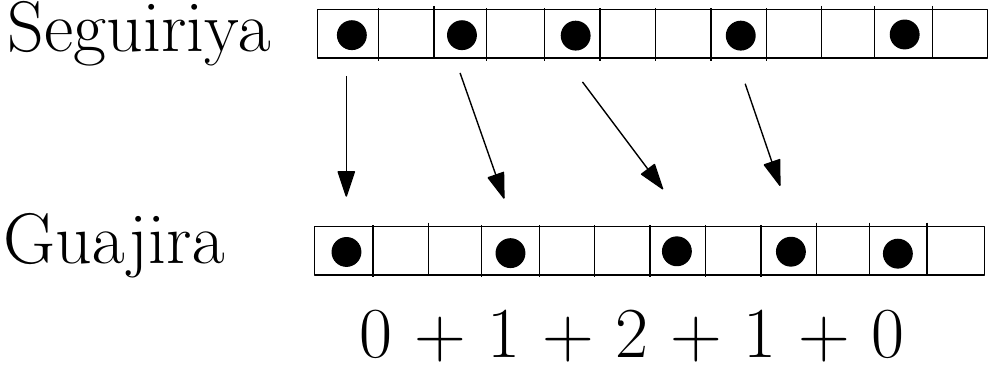}
 \caption{
               The black dots represent loud claps and the white dots are soft claps. %
   $d_p(Se,Gu)= |1-1 |+ |3-4 |+ |5-7 |+ |8-9 |+ |11-11|=4$.}
           \label{d_p(S,G)}
\end{center}
\end{figure}

In general, when we want to transform one string into another that does not have the same number of ones as the first string, the following restrictions apply (an example is shown in Figure \ref{d_p(S,F)}):
\begin{enumerate}
\item The rhythm with more strong accents $P_1$ (seguiriya in the example) is transformed to the rhythm with fewer strong accents $P_2$ (fandango in the example).
\item Each accent (loud clap) in $P_1$ must move to an accent in $P_2$.
\item Each accent in $P_2$ (fandango in the example) must receive at least one accent from $P_1$ (seguiriya).
\item The accents can not cross the end of the string and wrap around to the beginning.
\end{enumerate}

The general problem of calculating the permutation distance between any two long binary strings requires a more complex analysis which is beyond the scope of this work. We note that the assignment of accents for the case of strings with the same number of ones is a (one-to-one) bijection. However, for the case of strings with different numbers of ones, the assignment must be non-injective (several-to-one) or surjective. Readers who wish to know more about this topic may see \cite{colanino}.

\begin{figure} [htb]
                          \begin{center}
            \includegraphics[width=8cm]{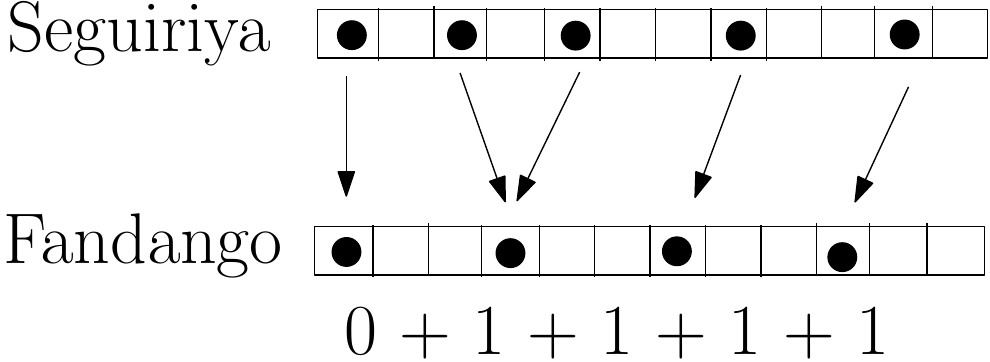}
 \caption{
                   The rhythm with more strong accents is transformed into the rhythm with fewer accents. %
               $d_P(Se, Fa)=4$.}
           \label{d_p(S,F)}
\end{center}
\end{figure}

Similarly to the way we compared compases using chronotonic distance, we can now obtain the permutation distance between all the ternary flamenco rhythms as shown in Table \ref{tabla-permutacion}. 


\begin{table}[h]
\begin{center}
\begin{tabular}{|l|c|c|c|c|c|}
\hline
$d_p$& Sole\'a & Buler\'ia & Seguiriya & Guajira & Fandango \\
\hline
\multirow{1}{12mm} {Sole\'a} &  0 & 1  & 11 & 7 & 7  \\
\hline
\multirow{1}{12mm} {Buler\'ia} &  1 & 0  & 12 & 8 & 8  \\
\hline
\multirow{1}{15mm} {Seguiriya} &  11 & 12  & 0 & 4 & 4  \\
\hline
\multirow{1}{12mm} {Guajira} &  7 & 8  & 4 & 0 & 2  \\
\hline
\multirow{1}{17mm} {Fandango} &  7 & 8  & 4 & 2 & 0  \\
\hline
\hline
\multirow{1}{12mm} {$\sum$} &  26 &  29  & \textbf{34} & \textbf{21} & \textbf{21} \\
\hline
\multirow{1}{12mm} {\emph{Max}} &  11 & \textbf{12} & \textbf{12} & \textbf{8} & 8  \\
\hline
\end{tabular}
\caption{
               Table of permutation distances.}
\label{tabla-permutacion}
\end{center}
\end{table}

We observe that for this similarity measure, when we take the sum of the distances, $\sum$, as the nearness criterion, the seguiriya is the most dissimilar to the others, and the guajira and fandango are tied as the nearest. 
A relatively small sum indicates the rhythm is very similar to all the others, whereas a relatively large sum means
it is distinct. In a musicological study, more similar to others can be interpreted as oldest in evolution.
 If we use the maximum distance, \emph{Max}, then the buler\'ia is again the most dissimilar comp\'as ($d_p(Bu,Se)=12$), as is the seguiriya\footnote{This casts doubt on the theory advanced in \cite{mairena-63} that the seguiriya is the primal flamenco comp\'as.}.

This similarity analysis could be applied in investigations of the origins of flamenco rhythms that interpret them as ``living beings'' that evolve over time under the influence of social parameters such as preferences and fashions. Following this metaphor, we can use bioinformatic techniques, putting the data into a similarity matrix in a \emph{phylogenetic tree} that can be drawn using tools such as SplitsTree\footnote{\url{http://www.splitstree.org}}. In Figures \ref{arbol-crono} and \ref{arbol-permutacion}, we have drawn the phylogenetic trees (in general, graphs are required) of the chronotonic and permutation distances respectively.
The ``living organisms'' (flamenco \emph{palos} in our case) are the nodes, and the distances between the nodes in the trees are the values given in the respective distance matrix. This methodology can be generalized to other musical parameters and used for a scientific analysis of the evolution and/or classification of flamenco styles\footnote{These tools should be used with caution and the interpretation of their results used merely to supplement analyses carried out from other scientific points of view such as historiography and anthropology.}.

\begin{figure} [htb]
                        \begin{minipage}[t]{5cm}
                          \begin{center}
            \fbox{\includegraphics[width=5cm]{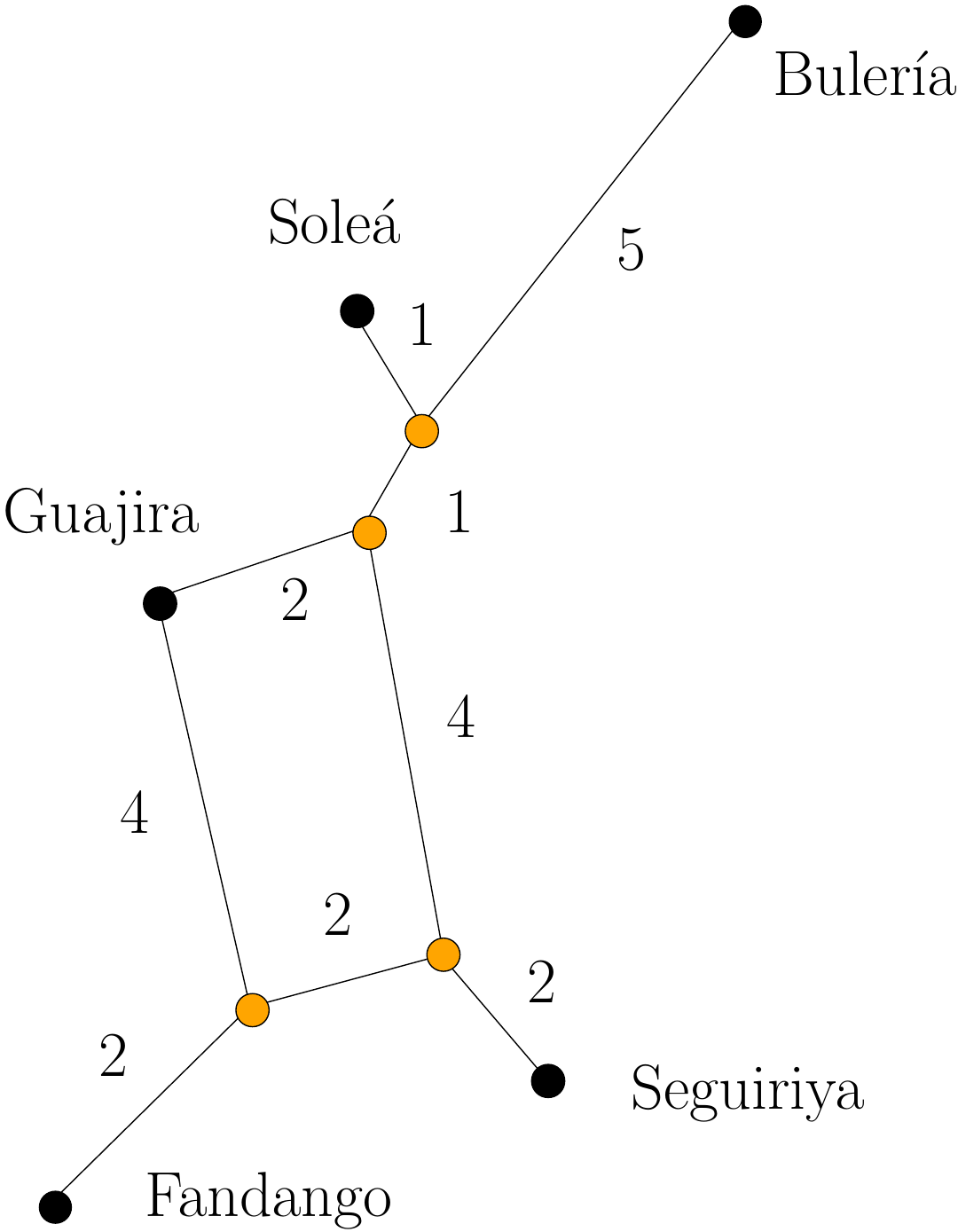}}
 \caption{
                Phylogenetic tree using chronotonic distance.}
           \label{arbol-crono}
\end{center}
                        \end{minipage}
                    \hfill
                        \begin{minipage}[t]{5cm}
                          \begin{center}
\fbox{\includegraphics[width=5cm]{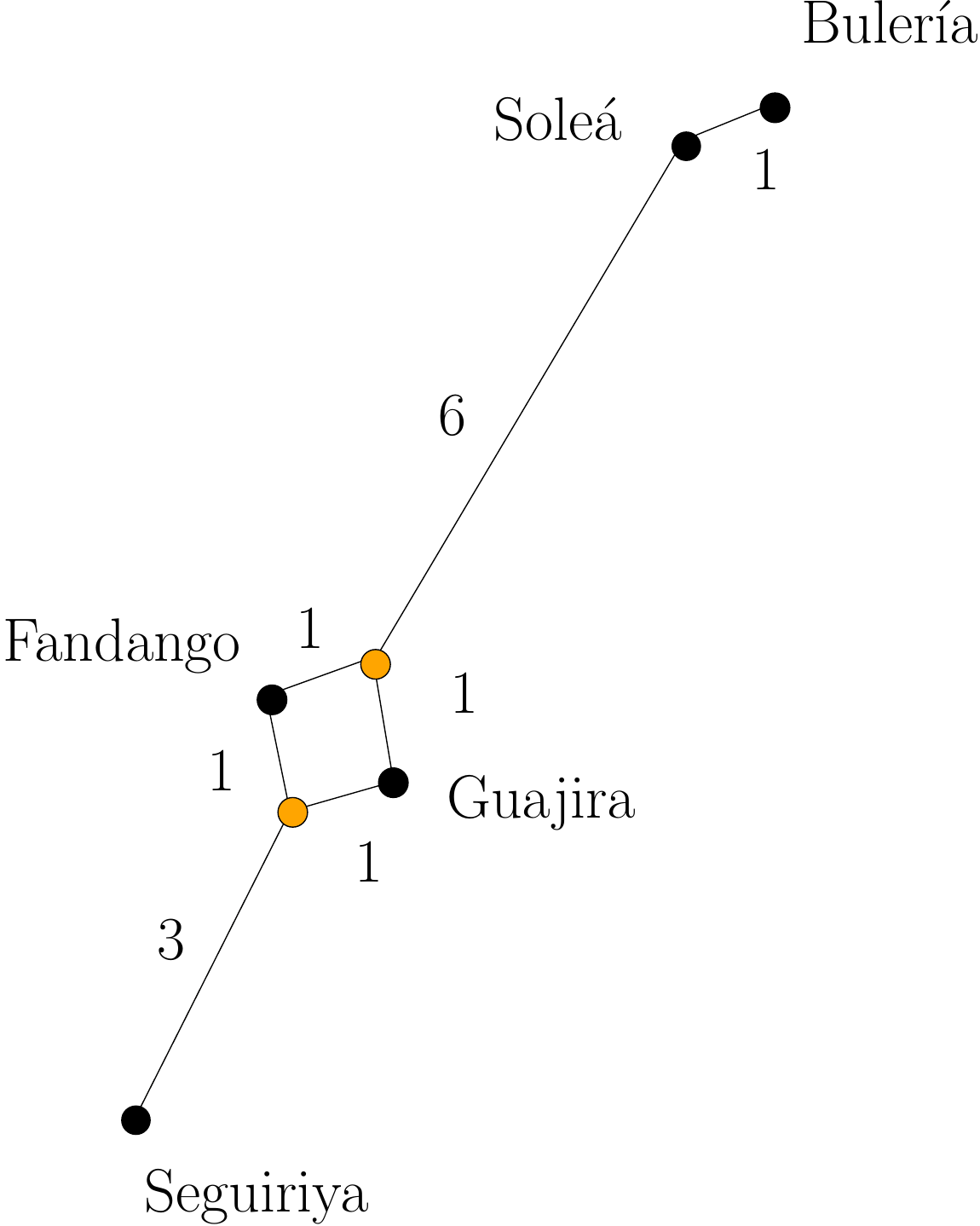}}
    \caption{
                  Phylogenetic tree using permutation distance.}
    \label{arbol-permutacion}
    \end{center}
                        \end{minipage}
                \end{figure}

\section{
              Approximating functions: Melodic simplification}

As we noted above, a melodic phrase extracted from a set of audio data can be a string made up of a large number of characters (pitches or notes). In flamenco music, many of the elements of these strings are the result of ornamentation or melismas, and only a few come from the underlying basic melody of the cante; that is, from the skeleton of notes that define the melody. In fact, a manual transcription or score is merely an approximation of the melody containing the few notes chosen as the representative notes that define the melody to the transcriber.

Manual transcription is expensive and subjective. If we forego this and use the audio signal, we must extract a discrete approximation of the continuous signal, representing the cante by a discrete series of notes or pitches. Figure \ref{smstools-debla} shows the audio signal of a debla and its approximation by a series of pitches. In this operation, we must deal with a fundamental concept in mathematics that has been extensively studied, as it has been used in a great number of real-world applications; namely, approximation theory.

The definitions of ``approximate'' in the Random House Dictionary are ``to approach closely, to bring near, to estimate.'' In mathematics, to approximate means to come as close as necessary to achieve a desired result.
In our context, the goal is to obtain results as close as possible with the purpose of working with functions that are simpler than the complicated exact functions.
A good approximation is  a representation that, although not exact, is sufficiently close to be useful for the problem at hand (for example, melodic similarity in flamenco cantes.)

\subsection{
                    Approximating a continuous function by a step function}

A step function is a piecewise constant function having only finitely many pieces. function in which the image of each element in the domain is a constant. Each of these constant values is called a step of the function. For example, the function $f(x)=k$ in each interval $[k-1, k)$ for $k=1,2,3, \ldots$ is a step function defined on the entire positive real line.

\begin{prob}[Similarity distance between step functions]
Given two step functions $f$ and $g$ that represent two melodies defined on the same time interval, design a measure of distance $\mathrm{Dist}(f,g)$ that represents the dissimilarity between the two melodies in some way. Two melodies with a small distance between them should correspond to similar songs.
\end{prob}

\begin{prob}[Approximation by a step function] \label{step-approximation}
Given a set of points $P = \{p_1, p_2, \dots , p_n\}$ on the plane and an error value $\alpha \geq 0$, construct a step function $E$ that approximates $P$ with error (defined as $\Max {d_v(p_i,E)}$) 
no greater than $\alpha$, using the minimum number of horizontal segments.
\end{prob}

The vertical distance of a point $p_i=(x_i,y_i)$ (where $x_i$ is in the domain of $f$) from a function $f$ is defined as $d_v(p_i,f)=|f(x_i)-y_i|$. The problem is thus one of designing an efficient procedure to solve this approximation problem.
In music, the interpretation of the problem is as follows. A melodic curve can be approximated by a step function. This is called \emph{segmentation} in music technology. Given that the range of the flamenco voice is normally limited to one octave, values of the error term $\alpha$ equal to a full or half tone should be suitable for extracting the melodic skeleton of a given cante. Thus the melody is encoded as a vector containing the steps of the approximation. These problems are related to the classification of flamenco cantes \cite{db-rizo}.  

\begin{figure}[!htb]
    \hspace*{-2cm}
   \includegraphics[width=1.3\textwidth]{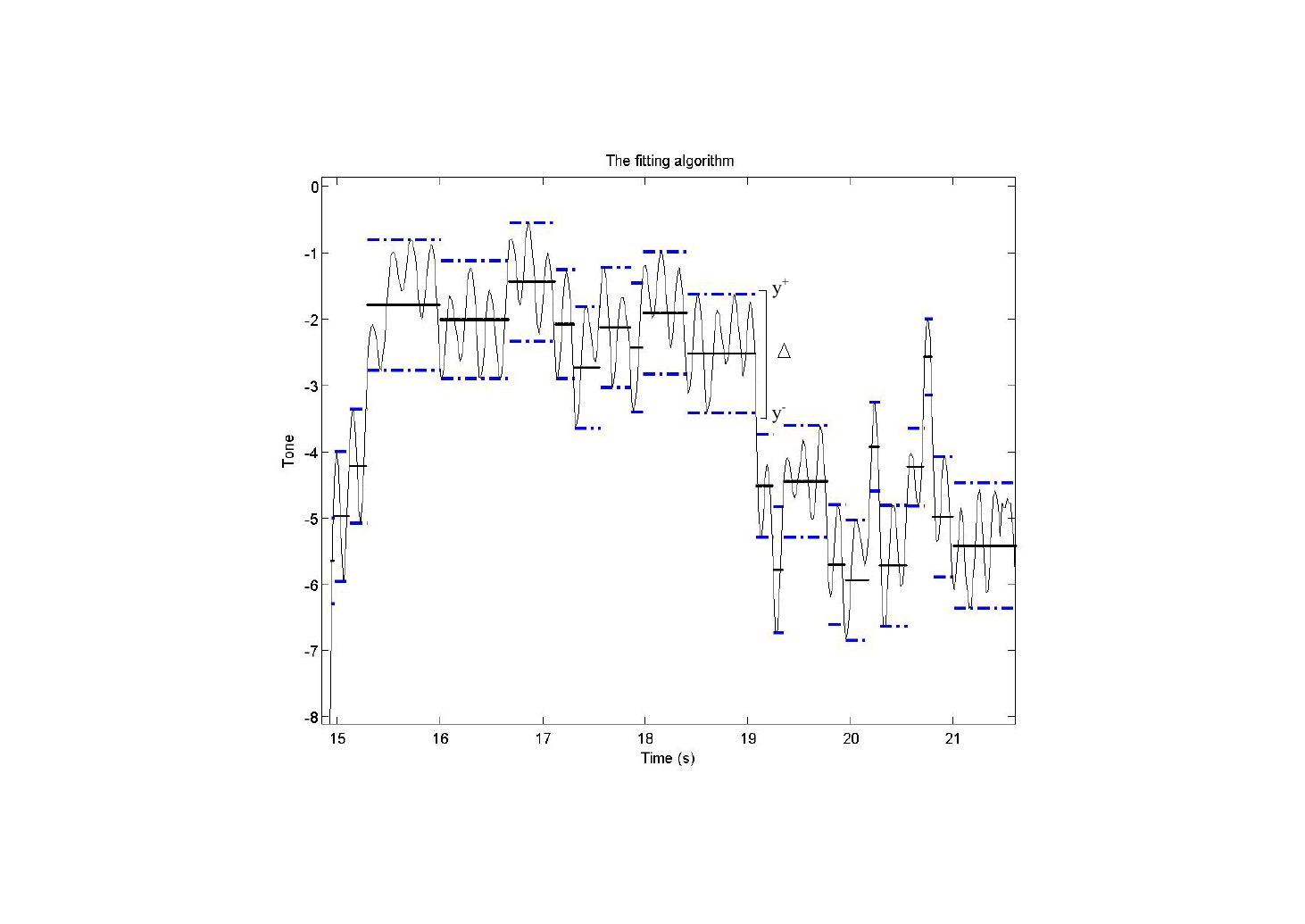}
   \vspace*{-2.5cm}
   \caption{
                 Construction of a step function to fit a flamenco melodic curve.}
   \label{fitting}
\end{figure}

We now describe a method given in \cite{dm01} that can be used to give an efficient solution to Problem \ref{step-approximation}. Given a set $P = \{p_1, p_2, \ldots , p_n\}$ of $n$ points and a tolerance error $\alpha$, let us draw vertical segments $V_i$ of height $2\alpha$ centered at each point $p_i$ as in Figure \ref{fitting}. The restriction that the vertical distance of each point $p_i$ from the value of the step function $E$ may be no more than $\alpha$ is the same as asking that $E$ intersects all the segments $V_i$ of length $2\alpha$. We proceed from left to right placing a step that intersects the greatest number of consecutive vertical segments possible, then another step, and so on. A vertical segment is defined as an interval of the variable $y$, say $[y_i^-, y_i^+]$, where $y_i^-$ and $y_i^+$ denote the coordinates of the low and high point respectively. Continuing from left to right, it is sufficient to maintain the intersection $\Delta$ with the vertical segments until reaching a segment $V_j$ that does not intersect with $\Delta$, in which case the current step ends and a new step is begun at $V_j$, updating $\Delta$ to  $ [y_j^-, y_j^+]$ (Figure \ref{fitting}). It is not difficult to prove that the number of operations is a linear function of $n$, the number of points in $P$, and the process results in a step function that fulfills the requirements of having a minimum number of steps and an error that does not exceed $\alpha$.

\begin{figure}[!htb]
   \centering
   \includegraphics[width=1.1\textwidth]{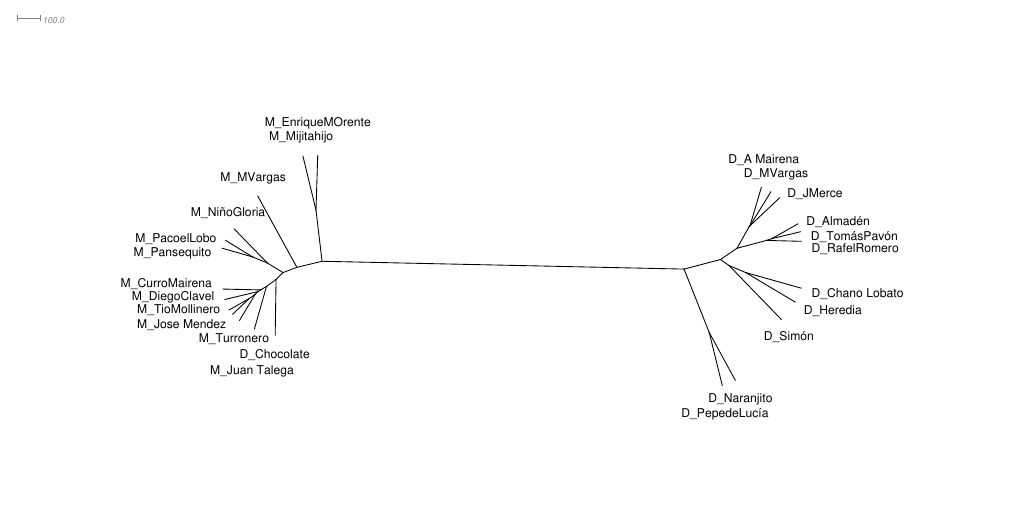}
   \caption{
                Before segmentation: well-defined groups, with one outlier; the debla by Chocolate.}
   \label{Tree-fitting1}
\end{figure}

\begin{figure}[!htb]
   \centering
   \includegraphics[width=1.15\textwidth]{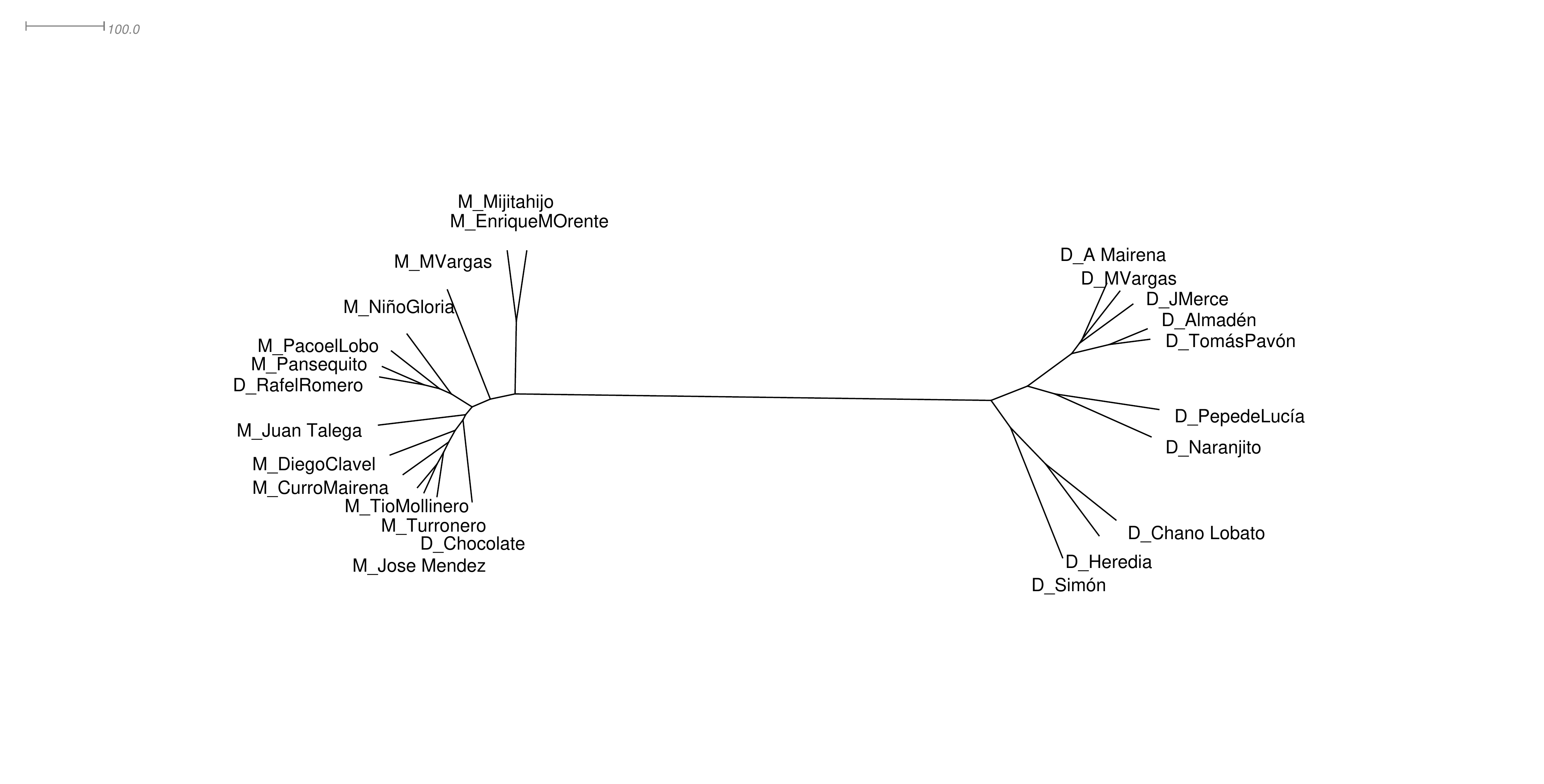}
   \caption{
                 After segmentation: well-defined groups, with two outliers: the deblas by Chocolate and Rafael Romero.}
   \label{Tree-fitting2}
\end{figure}

To find out whether approximation by step functions gives the desired result for flamenco music, we tested it on a corpus of flamenco cantes containing two styles, deblas and martinetes\footnote{The used dataset is a subset of the TONAS collection available at \url{www.cofla-project.com/data.html}.}, creating a phylogenetic tree from the similarity matrix of similarity distances. Details on the computation of the similarity matrix can be seen in \cite{db-rizo}. The segmentation algorithm spends less than
one second to compute the step functions of
the considered melodies collection and speeds up
both the time and space complexity of the standard calculation of the similarity matrix on the raw contours without segmentation.
The results are shown in Figures \ref{Tree-fitting1} and \ref{Tree-fitting2} where it can be seen that the clusters before and after simplification are similar, which validates the approximation method used.

\section{
              Some final observations}

This article has presented a small collection of problems that could be used in mathematics teaching and research. The singular aspect of this corpus of problems is its origin. The fact that these problems arose naturally in the study of flamenco, a folk art supposedly very different from mathematics, makes it a good candidate for introducing formal mathematical concepts in the classroom using appealing examples, such as the concept of permutation or swap (exchange) of characters in a string and the formal definition of distance (these would be suitable for teaching at the high school level), or dynamic programming and other algorithmic paradigms (in university courses).

The set of problems examined here could be extended to other relevant problems such as the design of search algorithms to recognize musical patterns, useful for classifying styles or variants. We describe an example that illustrates this problem: If we detect that a distinctive \emph{fandango de Valverde} pattern is present in a set of audio data, there is a high probability that it is a \emph{fandango de Huelva} cante. We refer the reader to the paper \cite{ismir2012-aggelos} for a recent study of this problem.
Other problems in music technology that require further scientific research include:
\begin{itemize}
\item Design a robust algorithm to separate out the voice part of a flamenco recording (most recordings do not separate guitar, percussion and voice) \cite{ismir2012-emilia}.
\item Design a method to estimate the pitch and segmentation of musical notes \cite{emilia-bonada}.
\item Design algorithms to automatically detect melodic motifs or ornamentation \cite{granada}.
\item Study the mathematical properties of preference in the sphere of flamenco music \cite{db-itamar}.
\end{itemize}

The main objective that leads us to develop mathematical and computational techniques such as those described in this article is to provide an objective tool to complement and contrast results with those from other scientific fields. Moreover, the multidisciplinary nature of research into flamenco music means that scientific progress can be made in research not only in mathematics, but in other areas as diverse as anthropology, engineering, education and musicology
 \cite{db-bridges}.

In conclusion, we note that in addition to the scientific and academic progress this would entail, it is not difficult to imagine potential commercial applications for the software that would emerge from the new music technology that would be developed in the course of such a research program -- music recommendation systems, on-line recognition of musical styles, and automatic classification of styles, to name a few examples.

\endgroup

\medskip \section*{
                Acknowledgments}

We would like to thank the Junta de  Andaluc\'ia (Consejer\'ia de Innovaci\'on, Ciencia y Empresas) under grants P09-TIC-4840 and P12-TIC-1362, and the FEDER funds of the European Union for supporting this research. Special thanks are also extended to Margaret Schroeder for her help in writing this paper.
An anonymous referee also
deserves our thanks for a careful reading of the manuscript and for many valuable
suggestions and comments.


\bibliographystyle{plain}

\end{document}